# Coexistence of spin frustration and spin unfrustration induced spontaneous exchange bias in Heusler alloys

J. Li[1], X. Wang[1], J.J. Deng[1], Y. Wang[1], L. Ma[1,*], F.X. Ma[1], D.W. Zhao[1], C.M. Zhen[1], D.L. Hou[1], E.K. Liu[2], W.H. Wang[2], and G.H. Wu[2]

[1]*Department of Physics, Hebei Advanced Thin Films Laboratory, Hebei Normal University, Shijiazhuang 050024, China*

[2]*Beijing National Laboratory for Condensed Matter Physics, Institute of Physics, Chinese Academy of Sciences, Beijing 100190, China*

**Abstract:**

The mechanism of spontaneous exchange bias (SEB) and the dominant factor of its blocking temperature are still unclear in Heusler alloys. Here, the related investigations are performed in $Mn_2Ni_{1.5}Al_{0.5}$ Heusler alloys with SEB. The results of both magnetic measurements and first-principles calculations confirmed that spin frustrated and unfrustrated antiferromagnetic (AFM) states coexist there and they have different magnetic anisotropies, which are essential for SEB. Based on a series of measurement strategies, we demonstrate that the frustrated AFM state undergoes a first-order magnetic transition to the superferromagnet (SFM) state with the help of an external magnetic field, and SFM is retained due to the first-order property of the magnetic transition. SEB originates from the interface coupling of multiple sublattices between the unfrustrated AFM state and SFM state. By analyzing the Arrott plot using the Landau model, we found that the internal field of the system dominates the blocking temperature of SEB, which paves the way for improving the blocking temperature.

* corresponding author. majimei@126.com

# I. INTRODUCTION

Exchange bias (EB) is one of the phenomena associated with the interfacial exchange interaction between antiferromagnetic (AFM) and ferromagnetic (FM) phases [1]. It is extensively used in various applications including ultrahigh-density magnetic recording, giant magnetoresistance and spin valve devices [2,3]. Recently, it is found that EB can also be obtained without magnetic field cooling, preparation under a magnetic field, or annealing in a magnetic field, which is called the spontaneous exchange bias (SEB) [4]. Therefore, the application of SEB would be more convenient. Some SEB materials have been rapidly developed, including Heusler alloys [4-10], spinels [11], perovskites [12,13] and anti-perovskites [14,15]. We have found that all of these SEB materials have a single phase of non-stoichiometric composition, which is quite different from the conventional EB (CEB) materials composed of composite phases. The blocking temperature ($T_B$) of CEB is usually determined by the Neel temperature of the AFM phase [1]. Hence, it is easy to increase $T_B$ by increasing the Neel temperature of AFM phase. In contrast, the factors that dominate $T_B$ of SEB are not known, and as far as we know, there is no work to report this problem. Meanwhile, the current work shows that $T_B$ of SEB materials other than $Mn_2PtGa$ and $Mn_{3.5}Co_{0.5}N$ usually does not exceed 50 K [4-15], which greatly limits the application of SEB. Mechanism can provide an effective solution to the problem in physical effect, thus the SEB mechanism is particularly important.

There are still some different perspectives on the SEB mechanism, which focus on two issues. One issue is whether the pinned phase is inherent or externally induced. Some works indicated that the pinned phase is inherent in the SEB system [5,6,11-15]. Other works pointed out that the pinned phase is induced by an external magnetic field [4,7,8-10]. The other issue is how the SEB is realized. Several authors suggested that SEB is caused by the interfacial coupling between the pinning and pinned phases [4-10], thus it can be considered to be a surface effect of the pinned phase like CEB. Others believed that SEB originates from the exchange interactions between different magnetic sublattices [11-15], so it can be considered to be a bulk effect different from CEB. However, at present the sufficient experimental evidence on the two issues is still lacked, thus the dominant factor of $T_B$ cannot be determined.

In this paper, the non-stoichiometric Mn-rich $Mn_2Ni_{1.5}Al_{0.5}$ with SEB was chosen to explore the mechanism of SEB and search for the dominant factor of $T_B$. Experiments and first-principles calculations confirmed that both the spin frustrated and unfrustrated AFM states consisted of Mn atoms occupying different magnetic lattices. The frustrated AFM state undergoes a first-order magnetic transition to the superferromagnet (SFM) state. SEB is derived from the interfacial coupling of multiple sublattices between the unfrustrated AFM state and SFM state. We also found that the dominant factor of $T_B$ was the internal field of the system.

## II. EXPERIMENTAL DETAILS

In this work, polycrystalline ingots of $Mn_2Ni_{1.5}Al_{0.5}$ were fabricated with Mn, Ni and Al of 99.9% purity using arc melting technique under argon atmosphere and 8 wt.% more Mn was added to avoid the loss of Mn during arc melting. $Mn_2Ni_{1.5}Al_{0.5}$ ribbons were prepared by melt spinning the as-cast alloys. The surface velocity of the copper wheel was 20 m/s. DC and AC magnetic measurements for $Mn_2Ni_{1.5}Al_{0.5}$ were performed using Physical Property Measurement System (PPMS, Quantum Design, Inc.) and the Magnetic Property Measurement System (MPMS, Quantum Design, Inc.).

The calculation of magnetic properties was performed by spin polarization treatment for spin frustrated and spin unfrustrated AFM state of $Mn_2Ni_{1.5}Al_{0.5}$ that executed by the Cambridge serial total energy package (CASTEP) code based on density functional theory (DFT) [16]. The exchange-correlation energy was represented by the Perdew-Wang generalized gradient approximation (GGA). The interaction between the nucleus and the valence electrons was described by the ultrasoft pseudopotential. The plane-wave energy cutoff was set to 450 eV and the 3×3×3 Monkhorst-Pack k-point sampling were used for the calculation of geometry optimization. The structures studied here were fully relaxed until energy and force were converged to $5\times10^{-5}$ eV/atom and 0.002 eV/Å, respectively.

## III. RESULTS AND DISCUSSION

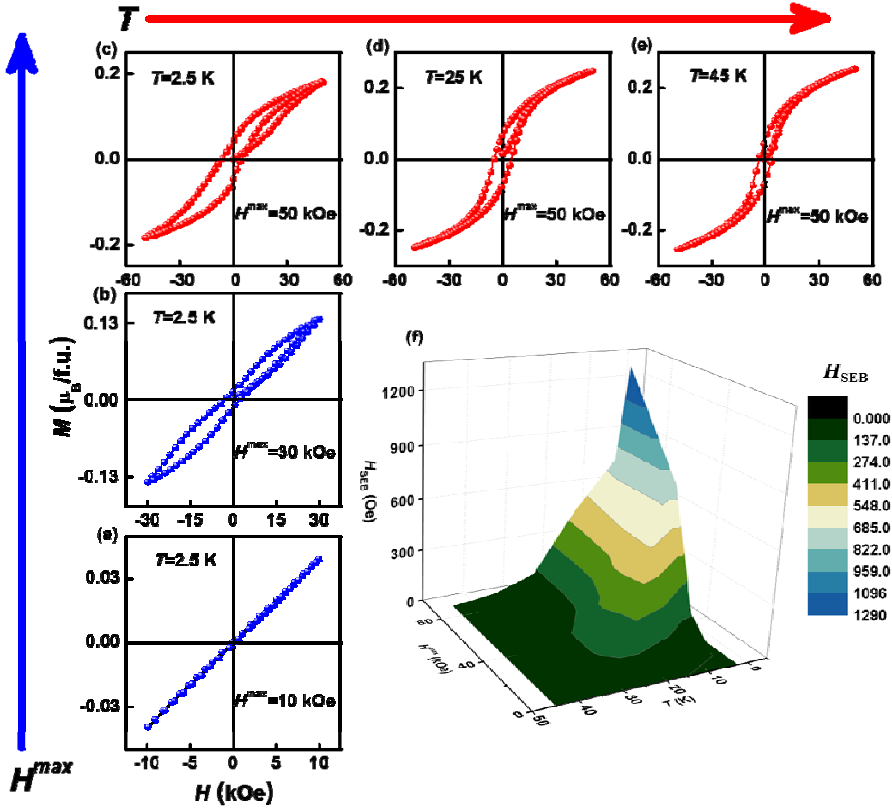

FIG. 1. $H^{max}$ and $T$ dependence of $H_{SEB}$. Selected virgin curves and hysteresis loops of $Mn_2Ni_{1.5}Al_{0.5}$ when increasing the maximum measurement magnetic field ($H^{max}$) at 2.5 K (a-c) and increasing the measurement temperature ($T$) under 50 kOe field (c-e), respectively. The blue and red arrows refer to the increase direction of $H^{max}$ and $T$, respectively. (f) $H^{max}$ and $T$ dependence of spontaneous exchange bias field ($H_{SEB}$) for $Mn_2Ni_{1.5}Al_{0.5}$.

Figure 1(a)-1(c) show the typical virgin curves and hysteresis loops of $Mn_2Ni_{1.5}Al_{0.5}$ at 2.5 K when increasing the maximum measurement magnetic field ($H^{max}$) from 10 kOe to 50 kOe. It is seen that the hysteresis loop under $H^{max}$ = 10 kOe shows a reversible straight line behaving like a single antiferromagnetic (AFM) phase. Under $H^{max}$ = 30 kOe, it becomes a double-shifted loop suggesting the coexistence of two magnetic orderings [17]. It is interesting that the loop shifts along the field axis

showing the spontaneous exchange bias (SEB) when $H^{max}$ is as large as 50 kOe, indicating the establishment of pinning [1]. Therefore, our measurement method well reflected that the same sample $Mn_2Ni_{1.5}Al_{0.5}$ has undergone complex transformations as $H^{max}$ increases. Figure 1(c)-1(e) show the typical virgin curves and hysteresis loops of $Mn_2Ni_{1.5}Al_{0.5}$ at $H^{max}$ = 50 kOe when increasing the measurement temperature ($T$) from 2.5 to 45 K. In the process of increasing $T$, we observe that the virgin curve moves from the outside to the inside of the hysteresis loop, which has also been observed in other intermetallic compounds and perovskite and was considered to be related to the field-induced magnetic transition [18,19]. In addition, all of the hysteresis loops with magnetic hysteresis are still rising for the high field part, indicating that FM and AFM orderings should coexist in the system. The above experimental results demonstrate that the temperature and magnetic field ($H$) together dominate the magnetism and SEB of the $Mn_2Ni_{1.5}Al_{0.5}$. Figure 1(f) shows the $T$ and $H$ dependence of the spontaneous exchange bias field ($H_{SEB}$). It is observed that $H_{SEB}$ shows the "cliff" change behavior along the $T$-axis or $H$-axis, rather than the "caret type" change behavior, just as the magnetic entropy change ($\Delta S$) varies with $T$ and $H$ due to the first-order magnetic transition [20].

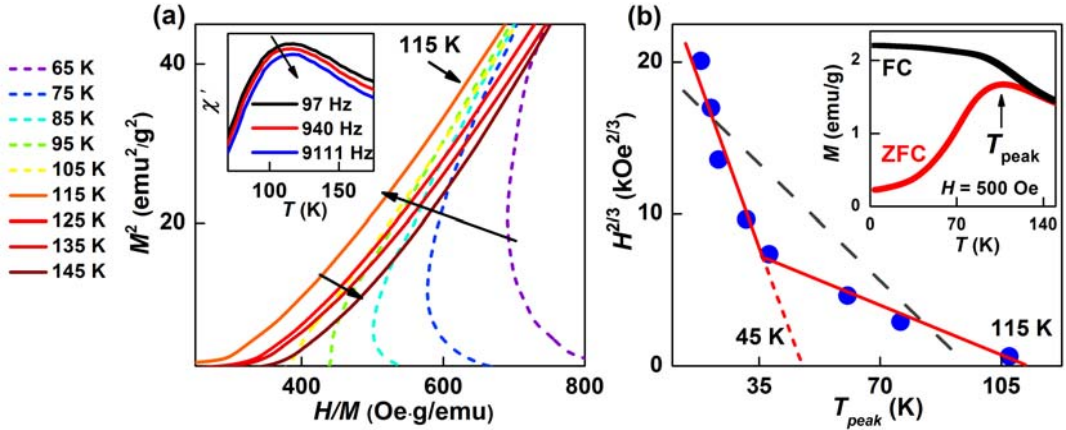

FIG. 2. The macroscopic evidence of the spin frustrated and unfrustrated AFM states. (a) Arrott plots for $Mn_2Ni_{1.5}Al_{0.5}$ at 65-145 K. The direction of temperature rise is indicated by the arrows. Inset shows the temperature dependence of the real part ($\chi'$) of AC susceptibility measured with an amplitude of 10 Oe at different frequencies. (b) Plot of $H^{2/3}$ vs $T_{peak}$ obtained from the zero-field cooling (ZFC) and field cooling (FC) curves curves under different external field ($H$). The solid

red lines are fit to the low and high field ranges, respectively, and the gray dashed line fits the entire field range, according to the Almeida-Thouless equation. Inset shows the typical ZFC and FC curves under $H$ = 500 Oe, the arrow denotes the temperature of the peak position ($T_{peak}$).

The inset of Figure 2(a) shows the real part ($\chi'$) of AC susceptibility as a function of temperature for $Mn_2Ni_{1.5}Al_{0.5}$. It is observed that as the frequency increases in the $\chi'$ curve, the amplitude of the peak gradually decreases, and the peak position at about 115 K moves toward the high temperature, indicating that there is a spin frustrated magnetic ordering in $Mn_2Ni_{1.5}Al_{0.5}$ [21]. Figure 2(a) shows the Arrott plots from 65 K to 145 K of $Mn_2Ni_{1.5}Al_{0.5}$. It is seen that the Arrott plots first move to the left before 115 K and then to the right after 115 K as the temperature increases, indicating that there is a strong AFM ordering in $Mn_2Ni_{1.5}Al_{0.5}$ [22,23]. As a consequence, the system exhibits the spin frustrated AFM state at 115 K.

Spin frustrated system follows the Almeida-Thouless (A-T) line behavior ($H^{2/3} \propto T_{peak}$) [21]. Hence, the zero-field cooling (ZFC) and field cooling (FC) curves at different external magnetic fields ($H$) from 0.5 kOe to 90 kOe were measured. Typical ZFC and FC curves at $H$ = 0.5 kOe is shown in the inset of Figure 2(b), and the arrow denotes the temperature of the peak position ($T_{peak}$). Figure 2(b) shows the corresponding plot of $H^{2/3}$ vs $T_{peak}$. However, for the entire temperature range, the plot does not obey the A-T line, as indicated by the gray dashed line. While the data in the low and high temperature ranges follow the A-T line well, as indicated by the two red dashed lines, two extrapolated characteristic temperatures 45 K and 115 K are obtained, where 115 K corresponds to the characteristic temperature of the spin frustrated AFM state. Accordingly, $Mn_2Ni_{1.5}Al_{0.5}$ is not a simple spin frustrated system [21]. It is worth noting that the system is highly sensitive to $H$ between 45 K and 115 K, the corresponding slope is -0.63 $kOe^{2/3}$/K, indicating that its anisotropy is weak, satisfying the spin frustrated characteristic. However, the dependence of the system on $H$ is insensitive between 0 K and 45 K, the corresponding slope is -10.96 $kOe^{2/3}$/K, indicating that it has a strong anisotropy in this temperature region, displaying the spin unfrustrated characteristic. In addition, the two negative slopes reflect the AFM ordering. Therefore, the spin frustrated and unfrustrated AFM states coexist in Mn$_2$Ni$_{1.5}$Al$_{0.5}$.

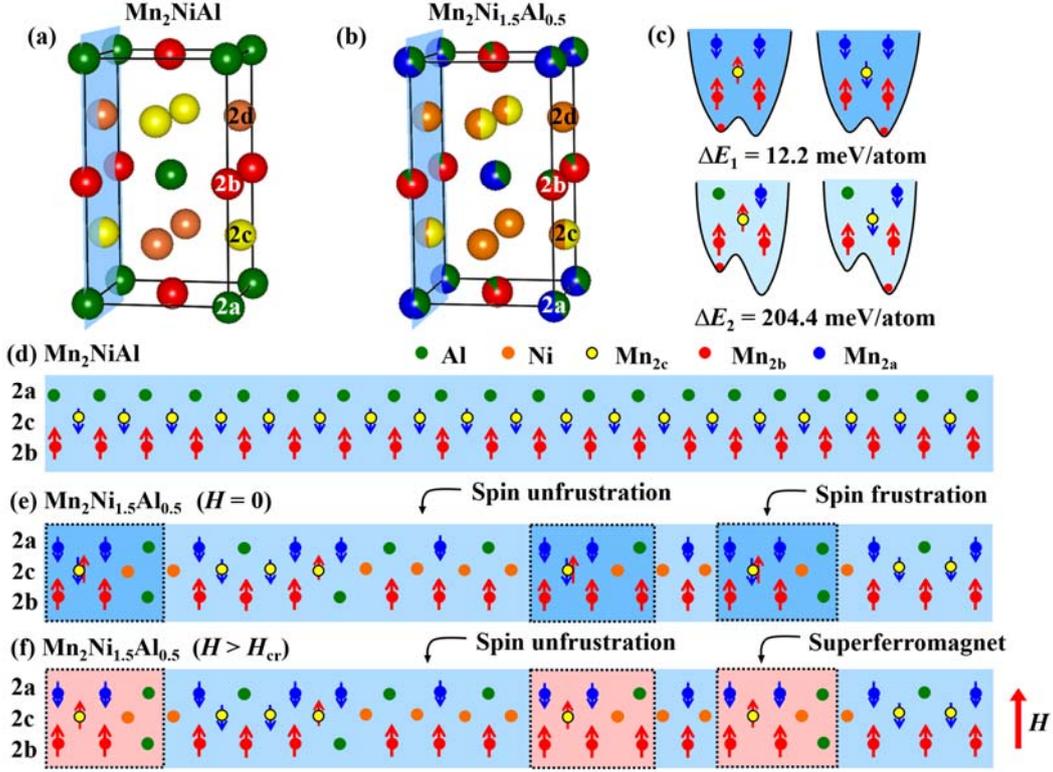

FIG. 3. Microscopic composition of the spin frustrated and unfrustrated AFM states. (a) Crystal structures of stoichiometric Mn$_2$NiAl and (b) non-stoichiometric Mn$_2$Ni$_{1.5}$Al$_{0.5}$, the (010) lattice planes are represented by the blue planes. Corresponding schematic diagram of the magnetic configuration on the (010) lattice plane for Mn$_2$NiAl (d) and Mn$_2$Ni$_{1.5}$Al$_{0.5}$ (e-f) at different external magnetic field respectively. Mn$_{2c}$, Mn$_{2b}$, and Mn$_{2a}$ represent that Mn atoms occupy the 2c, 2b, and 2a sites, respectively. (c) Energy difference of the system for the dark blue area $\Delta E_1$ and the light blue area $\Delta E_2$ when Mn$_{2c}$ is in different spin polarization directions.

In the Mn-based Heusler alloys, Mn is the main carrier of magnetic moment, its content and distribution determine the magnetic structure of the system. In order to understand the distribution and magnetic configuration of Mn atoms in non-stoichiometric Mn$_2$Ni$_{1.5}$Al$_{0.5}$, we first give the crystal structure of the stoichiometric Mn$_2$NiAl [space group $I\bar{4}m2$, see Figure 3(a)] in which four Mn atoms occupy 2c (0, 0, 0.5) and 2b (0, 0.5, 0.25) Wyckoff positions, denoted as Mn$_{2c}$ and Mn$_{2b}$, while two Ni atoms and two Al atoms occupy 2d (0, 0.5, 0.75) and 2a (0, 0,

0) Wyckoff positions, respectively. The corresponding magnetic configuration on the (010) lattice plane is shown in Figure 3(d). As can been seen, Mn atoms only occupy the 2c and 2b sites, and their magnetic moment direction are opposite [24,25]. In non-stoichiometric conditions, half of $Mn_{2c}$ atoms in $Mn_2Ni_{1.5}Al_{0.5}$ will occupy the 2a site of Al and become $Mn_{2a}$ [26,27]. In addition, part of $Mn_{2b}$ atoms will also occupy the 2a site owing to the anti-site disorder between $Mn_{2b}$ and Al atoms [25]. Thus, Mn atoms simultaneously occupy the 2c, 2b and 2a sites, and the number of $Mn_{2c}$ atoms is the smallest, as shown in Figure 3(b).

Figure 3(e) shows the corresponding magnetic configuration on the (010) lattice plane for $Mn_2Ni_{1.5}Al_{0.5}$. As can be seen, there are two kinds of neighboring environments for the $Mn_{2c}$ atoms, shown by the light and dark blue areas, respectively. We then perform first-principles calculations for these two areas, respectively. Figure 3(c) shows the calculation results. For the dark blue area, the energy difference of the system between the downward and upward spin polarization of $Mn_{2c}$ is 12.2 meV/atom, denoted as $\Delta E_1$. For the light blue area, the energy difference is 204.4 meV/atom, denoted as $\Delta E_2$, which is much larger than $\Delta E_1$, demonstrating that $Mn_{2c}$ is more stable in this area. Therefore, the dark blue area should represent the spin frustrated AFM state, the light blue area should represent the spin unfrustrated AFM state. In addition, it is worth noting that $\Delta E_1$ and the energy required for martensitic transformation are on the same order of magnitude [28].

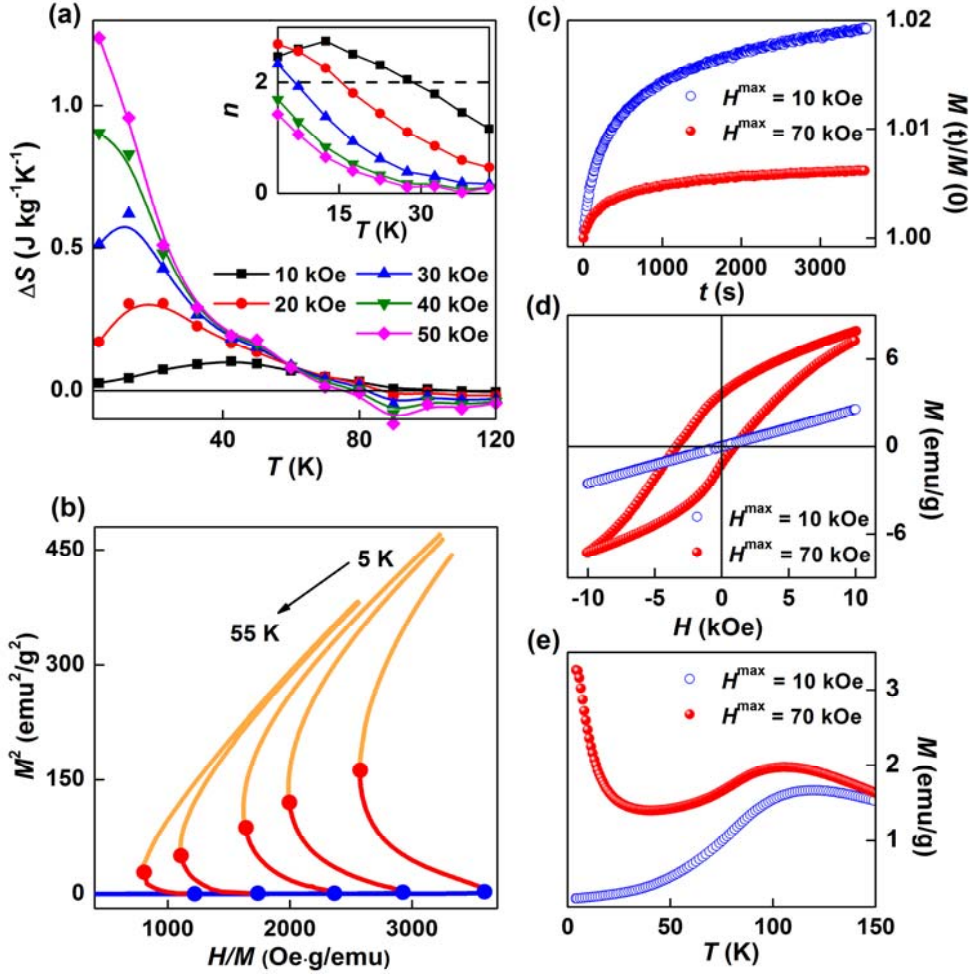

FIG. 4. Evidence of the first-order magnetic transition in the spin frustrated AFM state. (a) Entropy change ($\Delta S$) as a function of temperature and magnetic field for $Mn_2Ni_{1.5}Al_{0.5}$, the inset shows the magnetic field and temperature dependence of the exponent $n$ for $Mn_2Ni_{1.5}Al_{0.5}$. (b) Arrott plots for $Mn_2Ni_{1.5}Al_{0.5}$ from 5 K to 55 K. Solid blue and red circles represent the slope transition points. The blue, red and yellow portions represent that the curves have zero, negative and positive slopes, respectively. The dependence of the magnetization on time $M(t)$ under 0.5 kOe at 5 K (c), external field $M(H)$ at 5 K (d) and temperature $M(T)$ under 0.5 kOe (e) before (blue curves) and after (red curves) the first-order magnetic transition, respectively. The protocol before measurement (c-e) is that the $Mn_2Ni_{1.5}Al_{0.5}$ was first cooled down to 5 K under zero field, then subjected to a maximum initial field $H^{max}$ = 10 kOe and $H^{max}$ = 70 kOe, respectively, and finally AC demagnetization was performed at 5K.

Figure 4(a) shows the entropy change ($\Delta S$) as a function of the temperature and magnetic field for $Mn_2Ni_{1.5}Al_{0.5}$. It is worth mentioning that the sign of $\Delta S$ can qualitatively determine the order of phase transitions [29,30], especially $n(T,H) = \frac{dIn|\Delta S|}{dInH}$ is a quantitative criterion [20]. It is seen that Mn$_2$Ni$_{1.5}$Al$_{0.5}$ shows an obvious inverse magnetocaloric effect ($\Delta S > 0$), indicating that the magnetic transition is first-order and the lattice entropy change is larger than the magnetic entropy change [29,30]. The inset of Figure 4(a) shows the magnetic field and temperature dependence of the exponent $n$ calculated in Mn$_2$Ni$_{1.5}$Al$_{0.5}$ based on references [20,31]. It is found that the values of $n$ for $H \leq 30$ kOe are larger than 2, confirming that the first-order magnetic transition (FMT) occurs [20]. Besides, Arrott plots is also a generalized approach to distinguish first and second order magnetic transition, which is called Banerjee criterion [32]. Figure 4(b) shows the Arrott plots for Mn$_2$Ni$_{1.5}$Al$_{0.5}$ from 5 K to 55 K. It is seen that each curve has zero, negative and positive slopes represented by blue, red and yellow, respectively. The negative slope in Arrott plots combined with Banerjee criterion confirmed that FMT occurs in Mn$_2$Ni$_{1.5}$Al$_{0.5}$.

We designed a series of measurement strategies to compare the changes in magnetism before and after FMT in Mn$_2$Ni$_{1.5}$Al$_{0.5}$. The protocol before measurement is that the Mn$_2$Ni$_{1.5}$Al$_{0.5}$ was first cooled down to 5 K under zero field, then subjected to the maximum initial field $H^{max} = 10$ kOe and $H^{max} = 70$ kOe, respectively, to ensure that FMT does not occur and occurs, and finally AC demagnetization was performed at 5K. Figure 4(c)-4(e) show the $M(t)$, $M(H)$ and $M(T)$ before (blue curves) and after (red curves) FMT, respectively. As can be seen, before FMT, based on the blue curves the system exhibits obvious magnetic relaxation reflecting spin frustration [33], has a reversibly linear $M(H)$ demonstrating AFM behavior, and shows a peak at 115 K consistent with the results in Figure 2(a) and (b). Interestingly, after FMT, the system has undergone fundamental change in the above phenomena according to the red curves in Figure 4(c)-4(e). First, the magnetic relaxation is significantly suppressed indicating that the spin frustrated AFM state is destroyed. Thus, FMT occurs in the spin frustrated AFM state rather than the spin unfrustrated AFM state, which is agree with our first-principles calculation results. Second, both hysteresis and SEB appear in $M(H)$. It is thus proved that the spin frustrated AFM state has become SFM state with hysteresis, and the interfacial coupling between SFM state and the spin unfrustrated AFM state has been established. Third, by comparing the $M(T)$ before and after FMT as shown in Figure 4(e), SFM is rapidly destroyed as the temperature increases, since it is not inherent to the system but is retained due to FMT. Figure 3(e) and (f) show the schematic diagrams of magnetic configurations before and after FMT. The $Mn_{2c}$ atoms in the dark blue area is in a spin frustrated state, while the $Mn_{2c}$ atoms in the light blue area is in a spin unfrustrated state. When the external field is larger than the critical field ($H_{cr}$), the frustrated $Mn_{2c}$ moment points to the external magnetic field, FMT occurs and SFM is generated. SEB is derived from the interfacial coupling of multiple sublattices between the spin unfrustrated AFM state and SFM state.

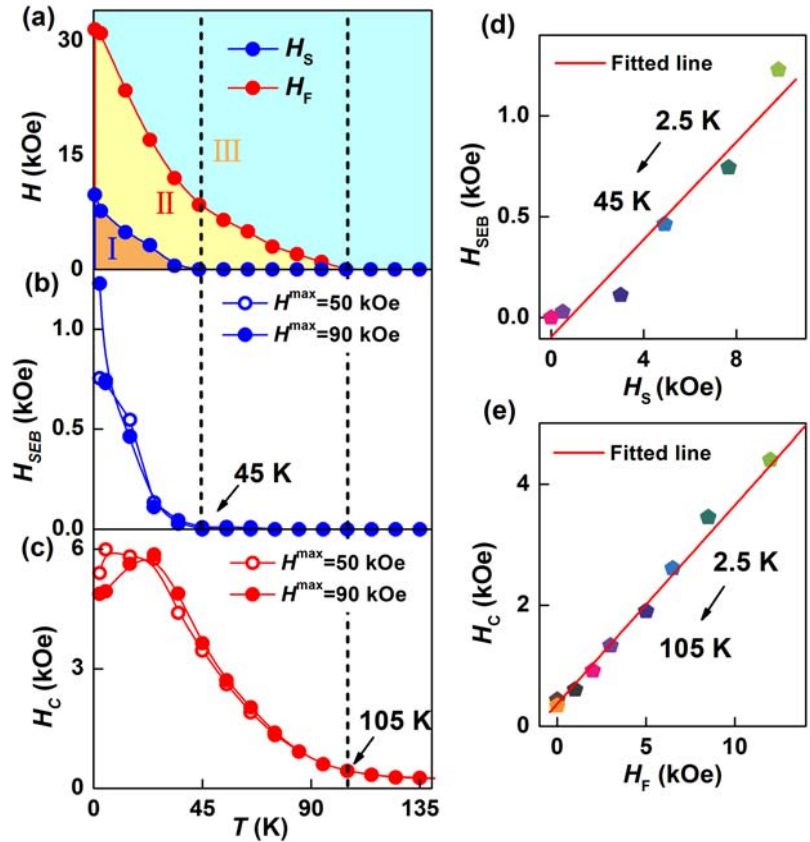

FIG. 5. The dominant factor of blocking temperature ($T_B$). (a) Temperature-magnetic field diagram of the first-order magnetic transition, $H_S$ and $H_F$ represent the start field and finish field of the first-order magnetic transition, respectively. Temperature dependence of the spontaneous exchange bias field $H_{SEB}$ (b) and the coercivity $H_C$ (c), taken from $M(H)$ curves under a maximum field of 50 kOe (open symbols) and 90 kOe (closed symbols), respectively. The vertical dashed lines are the disappearance temperatures of $H_S$ and $H_F$, respectively. Dependence of $H_{SEB}$ on $H_S$ (d), and $H_C$ on $H_F$ (e) for the Mn$_2$Ni$_{1.5}$Al$_{0.5}$. The increase in temperature is indicated by the black arrow.

Our above experimental and theoretical results confirmed that FMT and SEB are directly related. Therefore, there should be an intrinsic relationship between FMT and $T_B$. In this part, we will focus on this. Figure 5(a) shows the temperature dependence of the start field $H_S$ and the finish field $H_F$ of FMT, where the values of $H_S$ and $H_F$ are obtained from Figure 4(b). Figure 5(b) and (c) show the temperature dependence of $H_{SEB}$ and $H_C$ of SEB. As can be seen, there are striking similarities between $H_S$ and $H_{SEB}$, $H_F$ and $H_C$ from the temperature dependence. Accordingly, Figure 5(d) and (e) show the direct relationship between $H_S$ and $H_{SEB}$, $H_F$ and $H_C$. We found a linear correlation between $H_S$ and $H_{SEB}$, $H_F$ and $H_C$ by fitting, and the two red fitted lines passed the coordinate origin, i. e. when $H_S$ becomes zero, $H_{SEB}$ also becomes zero, $T_B$ reaches, and SEB disappears. Therefore, the dominant factor of $T_B$ is directly related to the physical meaning reflected by $H_S$. Based on the above experimental results, it can be concluded that increasing $H_S$ can effectively improve $T_B$. Therefore, understanding the physical meaning of $H_S$ is especially important.

Mn$_2$Ni$_{1.5}$Al$_{0.5}$ is made up of multiple magnetic sublattices. To simplify the situation, assume that the system consists of only two magnetic sublattices. According to the Landau model [22], the free energy of the system is written in the form: $F = \frac{1}{2}\widetilde{A}M^2 + \frac{1}{4}CM^4 - B_0 \cdot M + \frac{1}{2}\widetilde{a}m^2 + \frac{1}{4}cm^4 - B_0 \cdot m + \gamma M \cdot m$ Where $M$ and $m$ are the magnetic moments on each of the magnetic sublattice, $\gamma M \cdot m$ is the lowest order coupling term between $M$ and $m$, $\gamma$ is the coupling constant, the meaning of the other parameters and terms can be found in Ref. [22]. The influence of the terms $\gamma M$ and $\gamma m$ on the magnetic moments $m$ and $M$, respectively, is equivalent to a magnetic field, and the coupling terms proportional to $\gamma$ in the above equation have thus to be understood and interpreted in such a manner. It is the molecular field (Weiss field) of one sublattice acting on the sites of the other sublattice, and is defined as an internal field. Based on the discussion in Ref. [22], $\gamma$ in $Mn_2Ni_{1.5}Al_{0.5}$ is positive and has sufficient size, i. e. $M$ and $m$ are antiparallel and have strong coupling. More importantly, the Ref. [22] pointed out that the Arrott plot can reflect the competition between the internal field and the external field. The $H_S$ and $H_F$ of the Arrott plot divide Figure 5(a) into three regions denoted by I, II and III. In region I, the internal field is much stronger than the external one, and thus it can maintain the state of the antiparallel moment. At this moment, the net moment of the system remains at zero, which is reflected by the $M(H)$ in Figure 1(a). For large external fields (region III) the moments are from antiparallel to parallel. This can be seen from the (almost straight) lines of the Arrott plots for high fields in Figure 4(b). The regions of dominating internal and external magnetic fields are separated by a transition region [II in Figure 5(a)] where the spin frustrated AFM state becomes unstable for large fields, and the system switches in a first order transition to SFM state. $H_S$ is the start field of FMT, thus it can macroscopically reflect the strength of the internal field.

In this context, the direct relationship between $T_B$ and the internal field is constructed. At present, only $Mn_2PtGa$ (160 K) and $Mn_{3.5}Co_{0.5}N$ (256 K) have a $T_B$ of more than 150 K. For $Mn_2PtGa$, the strong spin-orbit coupling of Pt atom greatly improves the anisotropy of the system, which increases the internal field. For $Mn_{3.5}Co_{0.5}N$, the chiral AFM sublattice in the system due to the Dzyaloshinski-Moriya (D-M) interaction would also increase the internal field. Therefore, the way to improve $T_B$ is to increase the internal field.

## IV. CONCLUSION

In conclusion, $Mn_2Ni_{1.5}Al_{0.5}$ was selected to clarify the mechanism of spontaneous exchange bias (SEB) and the dominant factor of its blocking temperature ($T_B$). It is confirmed by magnetic measurements and first-principles calculations that $Mn_2Ni_{1.5}Al_{0.5}$ has both spin frustrated and spin unfrustrated antiferromagnetic (AFM) states. When the external magnetic field is higher than the critical field, the first-order magnetic transition occurs in the frustrated AFM state, and superferromagnet (SFM) state is induced. SEB is derived from the interface coupling between unfrustrated AFM state and SFM state on the sublattice scale. We found the direct relationship between the internal field of the system and $T_B$. Therefore, a method of increasing $T_B$ by increasing the internal field is proposed, which paves the way for improving the blocking temperature.

## ACKNOWLEDGEMENTS

This work was supported by the National Natural Science Foundation of China (Grant Nos. 11504247 and 11847017), the Hebei Natural Science Foundation (Grant Nos. A2014205051, E2016205268, and A2017210070), China Postdoctoral Science Foundation (Grant No. 2016M600192), the Department of Science and Technology of Hebei Province Scientific and Technological Research Project (Grant Nos. 13211032 and 15211036), the Science and Technology Research Project of Hebei Higher Education, China (Grant Nos. ZD2016042 and ZD2017045), the Natural Science Foundation of Hebei, China (Grant No. F2017208031).